\documentclass[aps,prb,twocolumn,groupedaddress,showpacs]{revtex4}

\usepackage{graphicx}

\begin{document}

\title{Connectivity transition in the frustrated $S=1$ chain revisited}

\author{A. K. Kolezhuk}
\affiliation{Institute of Magnetism, National Academy of Sciences and
Ministry of Education, 
36(b) Vernadskii avenue, 03142 Kiev, Ukraine }
\affiliation{Institut f\"{u}r Theoretische Physik,
Universit\"{a}t Hannover, Appelstr. 2, 30167 Hannover, Germany}
\homepage{http://www.itp.uni-hannover.de/~kolezhuk}

\author{U. Schollw\"{o}ck}
\affiliation{Sektion Physik, Ludwig-Maximilians-Universit\"at M\"{u}nchen,
Theresienstr.\ 37, 80333 Munich, Germany}

\date{October 3, 2001}

\begin{abstract}

The phase transition in the antiferromagnetic isotropic Heisenberg $S=1$
chain with frustrating next-nearest neighbor coupling $\alpha$ is
reconsidered. We identify the order parameter of the large-$\alpha$
phase as describing two intertwined strings, each possessing
a usual string order. The transition has a topological nature
determined by the change in the string connectivity. Numerical
evidence from the DMRG results is supported by the effective theory
based on soliton states.

\end{abstract}

\maketitle

In recent years, there has been a 
standing interest in
low-dimensional frustrated spin systems, motivated by their 
rich behavior which is often not completely understood.\cite{Diep94}
A fundamental example of such a system in one dimension is the
isotropic Heisenberg spin-$S$ chain with antiferromagnetic interactions
between nearest and next-nearest neighbors, described by the Hamiltonian
\begin{equation}
\widehat{H} = \sum_{i} {\mathbf S}_{i} {\mathbf S}_{i+1} +
\alpha \sum_{i} {\mathbf S}_{i} {\mathbf S}_{i+2}.
\label{ham}
\end{equation}
The physics of frustrated chains with half-integer $S$ is relatively
well understood. For small $\alpha$ they are gapless, and above the
certain $\alpha=\alpha_{c}$ a Kosterlitz-Thouless-type
transition into the gapped dimerized phase occurs, with
the exponentially slow opening of the gap.

\begin{figure}[tb]
\includegraphics[width=60mm]{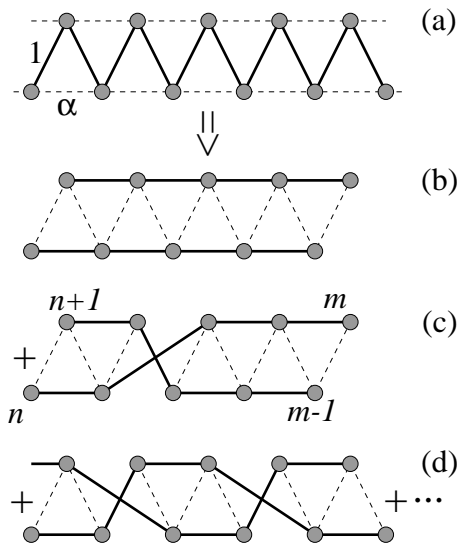}
\caption{\label{fig:string2} Schematic picture of the ground state of
the frustrated $S=1$ chain (\ref{ham}): (a) VBS state with one string
corresponding to the Haldane phase; (b)--(d) VBS states with two
strings contained in the DH (large-$\alpha$) phase.}
\end{figure}

For integer-$S$ chains the situation is much less clear.  At
$\alpha=0$ they are in the Haldane phase with a finite gap to the
elementary excitations.  For $S=1$, the Haldane phase
has the so-called \emph{string order} \cite{Nijs+89} arising due to the broken $Z_{2}\times Z_{2}$
symmetry\cite{Kennedy90} and  determined by the
nonlocal correlator
\begin{equation}
\label{sop1}
{\mathcal O}^{a}_{1}(n,n') = -\langle S^{a}_{n} 
( \exp \sum_{j=n+1}^{n'-1}i\pi S^{a}_{j}) 
S^{a}_{n'} \rangle,\;\; a=(x,y,z).
\end{equation}
The physics  of the $S=1$ chain is
believed to be well captured by the Affleck-Kennedy-Lieb-Tasaki (AKLT)
model\cite{AKLT} which differs from the Heisenberg model by the 
additional biquadratic exchange term ${1\over3}({\mathbf S}_{i}
{\mathbf S}_{i+1})^{2}$ in the Hamiltonian, and has the
exactly known ground state of the valence bond solid (VBS) type
 shown in Fig.\ \ref{fig:string2}a.

Several years ago it was observed\cite{KRS96} that the Haldane phase
in $S=1$ chain breaks down at $\alpha>\alpha_{c}\simeq0.75$ in a first
order transition, which is characterized by a discontinuous vanishing
of the \emph{string order parameter} (SOP) ${\mathcal
O}_{1}=\lim_{|n-n'|\to\infty}{\mathcal O}^{a}_{1}(n,n')$ at the transition
point, while the gap remains finite.  On the basis of simple energetic
arguments exploiting the AKLT-type variational states the transition
in the $S=1$ chain was heuristically interpreted \cite{KRS96} as a
decoupling of a single Haldane chain (cf. Fig.\ \ref{fig:string2}a)
into two subchains (cf. Fig.\ \ref{fig:string2}b). Later it was shown
\cite{Hikihara+00} that in the anisotropic chain the large-$\alpha$
``double Haldane'' (DH) phase persists in a finite region of
anisotropies.

However,  the  order parameter of the DH phase was never
identified, which made the real sense of the transition
rather unclear: complete decoupling would be reached only for
$\alpha\to\infty$, and at finite $\alpha$ the SOP
defined on a subchain is not the proper order parameter.
\cite{KRS96}

In the present paper we show,u sing the numerical results from the
density matrix renormalization group (DMRG) supported by the effective
theory based on soliton states, that the transition into the DH phase
corresponds to the decoupling into two \emph{intertwined substrings},
each having a usual string order, and identify the corresponding order
parameter.  The transition thus has a \emph{topological} nature
determined by the change in the string connectivity.

We start with recalling some basic 
facts about the AKLT model. Its exact ground state
can be represented in the
 compact \emph{matrix product}
(MP) form:\cite{Fannes+89,Klumper+91-93} 
\begin{eqnarray} 
\label{AKLT}
&& \Psi_{AKLT}=\mbox{tr}\{g_{1}g_{2}\ldots g_{N}\}, \\
&& g_{n}=(1/\sqrt{3})(\sigma^{+}|-\rangle_{n}+\sigma^{-}|+\rangle_{n}
-\sigma^{0}|0\rangle_{n}), \nonumber
\end{eqnarray}
where $\sigma^{\mu}$ are the Pauli matrices in the spherical basis,
$\mu=0,\pm1$, and $|\mu\rangle_{n}$ are the spin states at the $n$-th
site. The AKLT state possesses perfect string order, ${\mathcal
O}^{a}_{1}=4/9$. Elementary excitations of the AKLT chain are
\emph{solitons in the string order}.\cite{FathSolyom93} A soliton
with $S^{z}=\mu$ at the $n$-th
site  is well approximated by the MP state
\begin{equation} 
\label{crack} 
|\mu,n)=  \mbox{tr}\{g_{1}g_{2}\ldots g_{n-1}(g_{n}\sigma^{\mu}) 
g_{n+1}\ldots g_{N}\}
\end{equation}
It is more convenient to use the equivalent set of states $|r,n)$ with
$r=(x,y,z)$, defined by Eq.\ (\ref{crack}) with $\sigma^{\mu}$
replaced by the Pauli matrices in the Cartesian basis $\sigma_{r}$.  
Soliton states
$|r,n)$ with different $n$ are not orthogonal.
However, one can introduce the equivalent set of states
\begin{equation} 
\label{sol}
|r,n\rangle =\big\{ 3|r,n-1)+|r,n) \big\}/2\sqrt{2}
\end{equation}
which have the property $\langle r,n|
r',n'\rangle=\delta_{rr'}\delta_{nn'}$.  The state (\ref{sol}) can be
 represented in the same MP form (\ref{crack}) with
$(g_{n}\sigma^{\mu})$ replaced by the following matrix $f^{r}_{n}$:
\begin{equation} 
\label{Msol} 
f^{r}_{n}=\sqrt{2/3}\, \openone\, |r\rangle_{n}-(i/\sqrt{6})\,
\varepsilon_{rjl}\, \sigma_{j}\,|l\rangle_{n}.
\end{equation}
Here $|r\rangle_{n}$ are the Cartesian spin-1 states at the site $n$,
$(r,j,l)\in (x,y,z)$, and $\openone$ is the $2\times2$ unit matrix.
One can check that in
the soliton state $|r,n\rangle$ the string order correlators
${\mathcal O}^{r'}_{1}(l,l')$ with $r'\not=r$ change sign when 
$n$ gets inside the $(l,l')$ interval,
while  ${\mathcal O}^{r}_{1}(l,l')$ remains insensitive to
the presence of the soliton.

The ground state of the Heisenberg $S=1$ chain
differs from that of the AKLT model by the presence of a finite
density of soliton pairs.  Indeed, the action of the spin operator
$S_{n}^{\mu}$ on the AKLT state is to create the state
$|\mu,n)+|\mu,n-1)$. Thus generally the action of the Hamiltonian
produces states with soliton pairs of the type
$\sum_{r}|r,n; r,n'\rangle$, and  only in the special
case of the AKLT model the contributions of bilinear and biquadratic
terms cancel each other. One may say that the AKLT state
is a \emph{skeleton state} for the
Haldane chain, which gets dressed with the soliton pairs.

The variational energy of the VBS state of Fig.\ \ref{fig:string2}a
for the Hamiltonian (\ref{ham}) is $-4/3+2\alpha/9$ (per spin). It is
easy to see\cite{KRS96} that there is another VBS state, being a
product of two VBS strings defined on the $\alpha$-subchains, as shown
in Fig.\ \ref{fig:string2}b, whose energy $-4\alpha/3$ becomes lower
for $\alpha>3/4$. Again, the actual large-$\alpha$ ground state
differs from the skeleton state of Fig.\ \ref{fig:string2}b by
the finite density of soliton pairs. There are two types of
pairs, with the solitons sitting on the same $\alpha$-subchain and on
different subchains. Presence of pairs with solitons
residing on different $\alpha$-subchains destroys the
normal SOP defined on a single subchain. One can check 
that creation of such pairs
is equivalent to adding a finite admixture of states with intertwined
VBS strings shown in Fig.\ \ref{fig:string2}c,d.

There is, however, another order parameter surviving
pair creation. One can define \emph{double string
order}  as 
\begin{equation} 
\label{sop2} 
{\mathcal O}^{a}_{2}(n,m) = \langle S^{a}_{n}S^{a}_{n+1} 
( \exp \sum_{j=n+2}^{m-2}i\pi S^{a}_{j}) 
S^{a}_{m-1} S^{a}_{m} \rangle,
\end{equation}
and it is easy to check that it is not sensitive to
the presence of any type of soliton pairs. 
The corresponding order parameter ${\mathcal
O}_{2}=\lim_{|m-n|\to\infty}
{\mathcal
O}^{a}_{2}(n,m)$
 is finite in any
state which is a product of exactly \emph{two} VBS strings,
\emph{arbitrarily intertwined}, because one can always factorize (\ref{sop2})
into a product of two normal SOPs defined along those strings. On the
other hand, the correlator (\ref{sop2}) decays exponentially in
the AKLT state with only
\emph{one} VBS string, because in this case (\ref{sop2}) factorizes
into $\langle S^{a}_{n}S^{a}_{m}\rangle {\mathcal O}_{1}(m-1,n+1)$.  One may
say that ${\mathcal O}_{2}$ measures the \emph{connectivity} of the state,
telling us how many VBS strings are there.  It is natural to assume that
${\mathcal O}_{2}$ should be the order parameter characterizing the DH phase. 

\begin{figure}[tb]
\includegraphics[width=80mm]{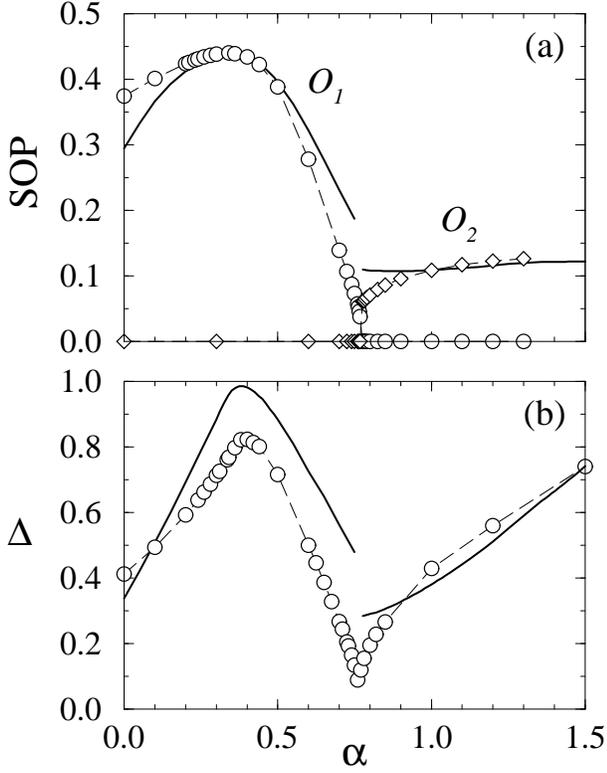}
\caption{\label{fig:result} (a) the string order ${\mathcal O}_{1}$ and
double string order ${\mathcal O}_{2}$ as functions of the frustration
$\alpha$; (b) the same for the excitation gap $\Delta$. Symbols
correspond to the DMRG data, and solid lines are theoretical
predictions from the effective model. }
\end{figure}

The above assumption is readily confirmed by the DMRG results. In our
DMRG calculations, we have considered chains with open boundary
conditions and up to $L=300$ spins, keeping $M=400$ states. We have
calculated the excitation gap and both the conventional SOP
 ${\mathcal O}_{1}$ and the new SOP ${\mathcal
O}_{2}$ (see Fig.\ \ref{fig:result}). 
While the results were converged in the number of states kept, we
carried out finite size extrapolation for frustrations very close to
the transition point.  We observe that both order parameters exclude
each other in the sense that if either of them decays to zero, the
thermodynamic limit of the other is non-zero and vice versa. The value
of ${\mathcal
O}_{2}$ is strictly zero below $\alpha<0.77$, and for
$\alpha=0.775$ already at 0.061, i.e. 44\% of its asymptotic
value of 0.1401 for $\alpha\rightarrow\infty$, where one has two
completely decoupled unfrustrated chains, and ${\mathcal O}_{2}$ is simply the
square of the string order for the unfrustrated $S=1$ Heisenberg chain,
0.3743. In fact, for the frustration $\alpha=1.3$, it reaches 90\%
of the asymptotic value.  At both sides of the transition,
both order parameters exhibit clearly finite values, signalling the
first order transition. The finding of a finite gap for all values of
$\alpha$ is consistent with this picture.

In order to describe the physical picture sketched above, we have
constructed the effective theory based on the soliton states
(\ref{sol}). For $\alpha<{3\over4}$ we regard the AKLT state 
as the vacuum state, and introduce  bosonic
operators $t_{nr}^{\dag}$ creating states
$|r,n\rangle$, $r=(x,y,z)$. 
Similarly, for $\alpha>{3\over4}$ we treat the two-subchain VBS
state shown in Fig.\ \ref{fig:string2}b as the vacuum; this state can
be conveniently represented in the following MP form:
\begin{equation} 
\label{AKLT2} 
\Psi_{AKLT2}=\mbox{tr}\big\{ \widetilde{\mathcal G}_{1} {\mathcal
G}_{2}\cdots  \widetilde{\mathcal G}_{N-1} {\mathcal
G}_{N} \big\},
\end{equation}
where ${\mathcal G}_{n}=(g_{n}\otimes \openone)$ and $\widetilde{\mathcal
G}_{n}=(\openone \otimes g_{n})$ are $4\times4$ matrices, and 
the total number of spins $N$ is assumed to be even.  The operators
$t_{nr}^{\dag}$ are interpreted as creating soliton states
$|\widetilde{r,n}\rangle$ on the subchains, defined by
replacing the $n$-th matrix ${\mathcal G}_{n}$  or $\widetilde{\mathcal G}_{n}$
in (\ref{AKLT2}) with the matrix ${\mathcal F}^{r}_{n}=(f^{r}_{n}\otimes
\openone)$ or  $\widetilde{\mathcal
F}^{r}_{n}=(\openone\otimes f^{r}_{n})$, respectively.

Calculating all matrix elements corresponding
to the hopping of solitons and creation of pairs, and passing to the
momentum representation, one obtains on the
quadratic level the
effective Hamiltonian of the form
\begin{equation} 
\label{Heff}
\widehat{H}_{eff} =\sum_{kr} A_{k}t^{\dag}_{k,r}t^{\phantom{\dag}}_{k,r}
+{1\over2}B_{k}(t^{\dag}_{k,r}t^{\dag}_{k,r}+\mbox{h.c.}).
\end{equation}
Here for the Haldane phase ($\alpha<{3\over4}$) the amplitudes
$A_{k}$, $B_{k}$ are given by the expressions
\begin{eqnarray} 
\label{AB-nn}
B_{k}&=&2(112\alpha/27-1)\cos{k} \nonumber\\
&-&(10\alpha/9)\cos{2k}
-(13\alpha-3)C^{(1,1)}_{k}, \nonumber\\
A_{k}&=&B_{k}+(4/27)\times\\
&\times&\big\{ 15-13\alpha +(9-16\alpha)\cos{k}
-3\alpha\cos{2k}\big\}, \nonumber
\end{eqnarray}
and for the DH phase
($\alpha>{3\over4}$) one has
\begin{eqnarray} 
\label{AB-nnn}
B_{k}&=&(2/9)(2\cos{k}-\alpha\cos{2k})
-C^{(3,2)}_{k}-\alpha C^{(4,2)}_{k}\nonumber\\
A_{k}&=&B_{k}+(4\alpha/9)(5+3\cos{2k}),
\end{eqnarray}
where the following shorthand notation is used:
\[
C^{(l',l)}_{k}={8\over9}\, 
{3\cos{(l'k)} +\cos{\big((l-l')k\big)} \over 5+3\cos{(lk)}}.
\]

The Hamiltonian (\ref{Heff})  does not take
into account any interaction between the solitons.  The most important
contribution to the interaction comes from the constraint that at most
one particle can be present at a given site, which is effectively
equivalent to the infinite on-site repulsion $U$.  We treat the
effect of the constraint using Brueckner's approximation along the
lines proposed by Kotov {\textit et al.}\cite{Kotov+98}. In this approach,
one neglects the contribution of anomalous Green's functions and
obtains in the limit $U\to\infty$ the vertex function
$\Gamma_{rr',ss'}=
\Gamma(k,\omega)(\delta_{rs}\delta_{r's'}+\delta_{rs'}\delta_{r's})$,
where $k$ and $\hbar\omega$ are respectively the total momentum and
energy of the incoming particles, with
\begin{equation} 
\label{Gamma}
{1\over\Gamma(k,w)} = -{1\over N}\sum_{q} 
{Z_{q}Z_{k-q}u^{2}_{q}u^{2}_{k-q}\over
\omega-\Omega_{q}-\Omega_{k-q}}.
\end{equation}
The corresponding normal self-energy $\Sigma(k,\omega)$ is
\begin{equation} 
\label{Sigma} 
\Sigma(k,\omega)=(4/N)
\sum_{q}Z_{q}v_{q}^{2}\Gamma(k+q,\omega-\Omega_{q}).
\end{equation}
The normal Green function has the form
\begin{equation} 
\label{Green}
G(k,\omega)={\omega+A_{k}+\Sigma(-k,-\omega) \over 
(\omega-\Sigma_{-})^{2} -(A_{k}+\Sigma_{+})^{2} +B_{k}^{2} } ,
\end{equation}
where $\Sigma_{\pm}\equiv{1\over2}\big\{\Sigma(k,\omega)
\pm \Sigma(-k,-\omega)\big\}$, and its 
quasiparticle part is given by
\begin{equation} 
\label{Green1}
G(k,\omega)={Z_{k}u_{k}^{2}\over \omega-\Omega_{k}+i\varepsilon} -
{Z_{k}v_{k}^{2}\over \omega+\Omega_{k}-i\varepsilon} 
\end{equation}
which defines the renormalization factors $Z_{k}$, the
Bogoliubov coefficients $u_{k}$, $v_{k}$ and the
spectrum $\Omega_{k}$ as follows:
\begin{eqnarray} 
\label{SO}
\Omega_{k}&=&\Sigma_{-}+E_{k},\quad 
E_{k}=\{ (A_{k}+\Sigma_{+})^{2}-B_{k}^{2} \}^{1/2},\nonumber\\
u_{k}^{2}&=&{1\over2}\big\{1+(A_{k}+\Sigma_{+})/E_{k}\big\},\quad
v_{k}^{2}=u_{k}^{2}-1,\nonumber\\
{1\over Z_{k}}&=&1-{\partial \Sigma_{-}\over \partial \omega} 
-{(A_{k}+\Sigma_{+})\over E_{k}}\,
{\partial \Sigma_{+}\over \partial \omega}
\end{eqnarray}
where $\Sigma_{\pm}$ and their derivatives are understood to be taken
at $\omega=\Omega_{k}$. The system of equations (\ref{Gamma}),
(\ref{Sigma}), (\ref{SO}) has to be solved self-consistently with
respect to $Z$ and $\Sigma$. This approach is valid
as long as the soliton density $\rho={3\over N}\sum_{q} Z_{q}v_{q}^{2}$
remains small, ensuring that the contribution of
anomalous Green's functions is irrelevant.\cite{Kotov+98}

It should be remarked that our Eqs.\ (\ref{SO}) differ from the
corresponding expressions of Kotov {\textit et al.,}\cite{Kotov+98} which
can be obtained from (\ref{SO}) assuming that $\Sigma(k,\omega)$ is
almost linear in $\omega$ in the frequency interval $(-\Omega_{k},
\Omega_{k})$; however, this latter assumption fails for the present
model.

As far as  Eqs.\ (\ref{Gamma}),
(\ref{Sigma}), (\ref{SO}) are solved, one can calculate the reduction
of the string order caused by the presence of pairs. For the usual SOP
in the Haldane phase ($\alpha<{3\over4}$) one finds
\begin{equation} 
\label{SOP1}
{\mathcal O}_{1}^{z}=(4/9)(1-2\gamma\rho_{p}R)^{2}.
\end{equation}
Here $\gamma={2\over3}$ is the
correcting factor which takes into account that $(zz)$ pairs do not
affect ${\mathcal O}_{1}^{z}$, and the total density of
soliton pairs $\rho_{p}={1\over2N}\sum_{nn'r}\langle
t^{\dag}_{nr}t^{\dag}_{n'r}\rangle^{2}$ and the mean size of the pair
$R$ are given by 
\begin{eqnarray} 
\label{rhop-mmean} 
&& \rho_{p}=(3/2N)\sum_{q} Z_{q}^{2} u_{q}^{2}v_{q}^{2},\nonumber\\
&& R={\sum_{m\geq 1} m w_{m}^{2}\over \sum_{m\geq1}
w_{m}^{2}},\quad w_{m}={1\over N}\sum Z_{q} u_{q}v_{q}e^{iqm}.
\end{eqnarray}

A similar calculation for ${\mathcal O}_{2}$ in the DH phase 
yields
\begin{equation} 
\label{SOP2} 
{\mathcal O}_{2}^{z}=(4/9)^{2}\big\{1- \gamma\rho_{p}^{l}R^{l}
\big\}^{4} \big\{1-2\gamma\rho_{p}^{t}R^{t} \big\}^{2},
\end{equation}
where $\rho_{p}^{l}$ and $\rho_{p}^{t}$ are densities of the
``longitudinal'' (sitting on a single subchain) and ``transversal''
(intersubchain) pairs, respectively, 
\begin{eqnarray} 
\label{rhos} 
\rho_{p}^{l,t}&=&(\rho_{p}\pm \widetilde{\rho}_{p})/2,\nonumber\\ 
\widetilde{\rho}_{p}&=&
{3\over 2N}\sum_{q}Z_{q}Z_{q+\pi}(u_{q}v_{q})(u_{q+\pi}v_{q+\pi}),
\end{eqnarray} 
and $R^{l}$, $R_{t}$ are the
corresponding average pair sizes (note that all distances are defined
in terms of the initial chain):
\begin{equation} 
\label{ms} 
R^{l}={ \sum_{m\geq 1} (2m) w_{2m}^{2} \over \sum_{m\geq 1}w_{2m}^{2}}, \quad
R^{t}={ \sum_{m\geq 0} (2m) w_{2m+1}^{2} 
\over \sum_{m\geq 0}w_{2m+1}^{2}}.
\end{equation}

In Fig.\ \ref{fig:result} we show the results of the DMRG calculations
together with the theoretical curves obtained on the basis of the
effective theory. Though the theoretical calculations
are not quantitatively satisfactory, they nevertheless capture the
essential behavior of the system. In the vicinity of the transition
the theoretical results are rather far from the numerical data; one
reason is that at $\alpha={3\over4}$ the completely dimerized state
has the same variational energy as the VBS states in Fig.\
\ref{fig:string2}(a,b), which is not taken into account in the theory.

It should be mentioned that a definition similar to (\ref{sop2}) was
recently used by Todo {\textit et al.}\cite{Todo+01} for the spin-1
ladder, with $(n,n+1)$ and $(m-1,m)$ placed on two rungs.
 They have shown that such an order is present in the $S=1$
ladder for any ratio of the rung and leg exchange. 
One may thus expect that in the frustrated $S=1$ chain with
alternating nearest-neighbor interaction the DH phase is smoothly
connected to the dimerized phase. However, this problem is not
 clear since in a ladder there are
\emph{three} ways to define the double string order, depending on
whether the initial and final pairs of points are placed on the rungs
or diagonals, and only one of the definitions gives a finite value
in the completely dimerized state.

In summary, we have identified the nature of the first order transition
in the frustrated $S=1$ chain as a change in the connectivity of
underlying VBS states, and established the proper order parameter for
the large-$\alpha$ phase. We have also developed the effective
description based on soliton states, which qualitatively describes the
physics of both phases. 

A.K. gratefully acknowledges the hospitality of the Institute for
Theoretical Physics, Hannover, where the present study was initiated.
This work is supported in part by the grant I/75895 from the
Volkswagen-Stiftung. U.Sch. is supported by a Gerhard-Hess prize of
the DFG.


\begin{thebibliography}{10}

\bibitem{Diep94} 
H. T. Diep (ed.), {\textit Magnetic
Systems with Competing Interactions (Frustrated Spin Systems)},
World
Scientific, Singapore, 1994.

\bibitem{Nijs+89} M. den Nijs and K. Rommelse, 
Phys.\ Rev.\ {\bf B 40}, 4709 (1989); S.M. Girvin, D.P. Arovas, 
Phys.\ Scr.\ T {\bf 27}, 156
(1989); T. Kennedy, H. Tasaki, Phys.\ Rev.\ {\bf 45}, 304 (1992).

\bibitem{Kennedy90} T. Kennedy, J.\ Phys.: 
Cond.\ Matter {\bf 2}, 5737 (1990).

\bibitem{AKLT} I. Affleck, T. Kennedy, E.H. Lieb, H. Tasaki, 
Phys.\
Rev.\ Lett.\ {\bf 59}, 799 (1987); 
Commun.\
Math.\ Phys.\ {\bf 115}, 477 (1988).

\bibitem{KRS96} A. Kolezhuk, R. Roth, and U. Schollw\"ock,  
Phys. Rev. Lett. {\bf 77}, 5142 (1996); Phys. Rev. B {\bf 55},
    8928 (1997).

\bibitem{Hikihara+00} T. Hikihara, M. Kaburagi, H. Kawamura, and
T. Tonegawa, J.\ Phys.\ Soc.\ Jpn.\ {\bf 69}, 259 (2000).

\bibitem{Fannes+89} M. Fannes, B. Nachtergaele and R. F. Werner,
Europhys. Lett. {\bf 10}, 633 (1989); Commun.\ Math.\ Phys.\ {\bf 144}, 443
(1992).

\bibitem{Klumper+91-93} A. Kl\"umper, A. Schadschneider and
J. Zittartz, J. Phys.  A {\bf 24}, L955 (1991); Z. Phys. B {\bf 87},
281 (1992); Europhys.\ Lett. {\bf 24}, 293 (1993).

\bibitem{FathSolyom93} G. F\'ath and J. S\'olyom, J. Phys.:
Cond. Matter {\bf 5}, 8983 (1993); 
U. Neugebauer and H.-J. Mikeska, Z. Phys. B {\bf 99},
151 (1996).

\bibitem{Kotov+98} V. N. Kotov, O. Sushkov, Zheng Weihong, and
J. Oitmaa, Phys. Rev. Lett. {\bf 80}, 5790 (1998).

\bibitem{Todo+01} S. Todo, M. Matsumoto, C. Yasuda, and Hajime
Takayama, \eprint{cond-mat/0106073}.

\end{thebibliography}
\end{document}